\documentclass[12pt,english]{article}

\usepackage[T1]{fontenc}
\usepackage[latin9]{inputenc}
\usepackage{geometry}
\usepackage{amsmath}
\usepackage{amsthm}
\usepackage{graphicx}
\usepackage{setspace}
\makeatletter
\numberwithin{equation}{section}
\numberwithin{figure}{section}
\newcommand{\lyxaddress}[1]{
	\par {\raggedright #1
	\vspace{1.4em}
	\noindent\par}
}

\makeatother

\usepackage{babel}
\begin{document}
\title{\textbf{On Mössbauer rotor effect, clock synchronization and third
postulate of relativity}}
\maketitle
\begin{center}
\textbf{\large{}Christian Corda}{\large\par}
\par\end{center}

\lyxaddress{\textbf{SUNY Polytechnic Institute, 13502 Utica, New York, USA; Istituto
Livi, 59100 Prato, Prato, Italy; International Institute for Applicable
Mathematics and Information Sciences, B. M. Birla Science Centre,
Adarsh Nagar, Hyderabad 500063, India. E-mail: }\textbf{\emph{cordac.galilei@gmail.com}}}
\begin{abstract}
The Mössbauer rotor effect recently gained a renewed interest due
to the discovery and explanation of an additional effect of clock
synchronization which has been missed for about 50 years, i.e. starting
from a famous book of Pauli, till some more recent experimental analyses.
The theoretical explanation of such an additional effect is due to
some recent papers in the general relativistic framework. Here we
show that the additional effect of clock synchronization can be calculated
in another way via the third postulate of relativity.
\end{abstract}
In the general relativistic framework, Einstein grasped that the gravitational
field can be represented via space-time curvature in a paper which
analysed the rotating frame, verbatim \cite{key-1}:

``\emph{The following important argument also speaks in favor of
a more relativistic interpretation. The centrifugal force which acts
under given conditions of a body is determined precisely by the same
natural constant that also gives its action in a gravitational field.
In fact we have no means to distinguish a centrifugal field from a
gravitational field. We thus always measure as the weight of the body
on the surface of the earth the superposed action of both fields,
named above, and we cannot separate their actions. In this manner
the point of view to interpret the rotating system K' as at rest,
and the centrifugal field as a gravitational field, gains justification
by all means. This interpretation is reminiscent of the original (more
special) relativity where the pondermotively acting force, upon an
electrically charged mass which moves in a magnetic field, is the
action of the electric field which is found at the location of the
mass as seen by the reference system at rest with the moving mass.}''

Such an interpretation enabled various general relativistic treatments
of Mössbauer rotor {[}2\textendash 10{]} and Sagnac effects \cite{key-11},
via the Einstein Equivalence Principle (EEP) which states the equivalence
between gravitation and inertia {[}4\textendash 11{]}. The EEP indeed
includes the rotating reference frame {[}4\textendash 10{]}. The discovery
and explanation of a new effect, neglected for over 50 years since
Pauli's famous book on relativity \cite{key-2}, has led to a recent
new interest in the Mössbauer rotor effect. The correct theoretical
explanation of this new effect, which is added to the traditionally
known one, is due to a series of works in which the analysis was carried
out with a general relativistic treatment {[}4\textendash 10{]}. In
this letter this additional effect of clock synchronization will be
calculated in a simple way via the third postulate of relativity.

This effect is due to the German physicist R. Mössbauer in 1958 \cite{key-12},
and involves gamma rays' resonant and recoil-free emission and absorption,
without loss of energy, by atomic nuclei bound in a solid. The particular
case of the \emph{Mössbauer rotor effect} (see Figure 1), shows an
absorber orbiting around the radiation's source. In Figure 1 it is
assumed that the $z-axis$ is perpendicular to the plane of the apparatus
within the circumference and that the same apparatus rotates around
such a $z-axis$ having constant angular velocity $\omega$. The final
detector in the right of the Figure is assumed at rest. The idea is
to measure the so-called \emph{transverse Doppler effect} via the
fractional energy shift for the resonant absorber {[}3\textendash 10{]},
because the transverse Doppler effect involves a relative energy shift
between emission and absorption lines through its motion. 
\begin{figure}
\includegraphics{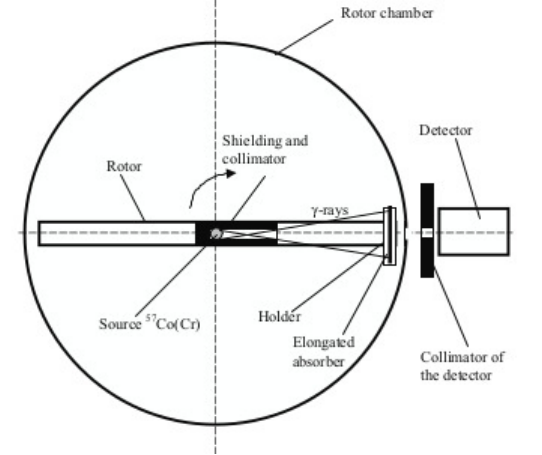}
\begin{description}
\item [{Figure}] \textbf{1} The scheme of the Mössbauer rotor experiment
in this Figure is adapted from \cite{key-14}. One assumes that the
$z-axis$ is perpendicular to the plane of the Figure and that the
apparatus within the circumference rotates around such a $z-axis$
having constant angular velocity $\omega$. The final detector in
the right of the Figure results at rest. 
\end{description}
\end{figure}

Kündig developed a fundamental Mössbauer rotor experiment in 1963
\cite{key-3}. The shift of the 14.4-keV Mössbauer absorption line
of $Fe^{57}$ was measured in function of the angular velocity of
the rotor \cite{key-3}. The result appeared to be consistent with
the theory of relativity with a sensitivity of $1.1\%$. Kündig also
discussed potential systematic errors \cite{key-3}. Kündig's experiment
has been recently investigated by an experimental group \cite{key-13,key-14}.
In the first work \cite{key-13} , the original data of Kündig's paper
have been reanalysed. Because of the strangeness of the results in
\cite{key-13}, the cited experimental group realized an independent
Mössbauer rotor experiment \cite{key-14}. In \cite{key-13}, it was
shown that mistakes were present in Kündig's original data. After
correcting those mistakes, the Authors found a higher value of the
fractional energy shift \cite{key-13}

\begin{equation}
\frac{\bigtriangleup E}{E}=-k\frac{v^{2}}{c^{2}},\label{eq: k}
\end{equation}
with $k=0.596\pm0.006$ where Kündig \cite{key-3}, Pauli \cite{key-2}
and others found $k=0.5$ in agreement with the relativity theory.
This was a strange issue, but here it will be shown that a correct
relativistic analysis can solve it. We recall that the Authors of
\cite{key-13} found that: i) the deviation of the coefficient $k$
Eq. (\ref{eq: k}) from $0.5$ is much higher than the measuring error
(about 20 times); ii) that deviation did not arise from some kinds
of disturbing factors like rotor vibrations because the excellent
methodology applied by Kündig \cite{key-3} surely excluded those
potential disturbing factors. In \cite{key-13} it was also emphasized
that also the experiment in \cite{key-15} appeared to be consistent
with $k>0.5.$ Remarkably, in the experiment in \cite{key-15} much
more data than similar Mössbauer rotor experiments analysed in {[}16\textendash 19{]}
have been involved. Based on the strange results in \cite{key-13},
the experimental group decided to realize a new experiment \cite{key-14},
where neither the scheme of Kündig's experiment \cite{key-3}, nor
the schemes of the other previously cited experiments {[}15\textendash 19{]}
have been followed. This permitted them to get independent information,
with respect to previous experiments, about the real value of $k$
in Eq. (\ref{eq: k}). The final result of the value of $k$ was $k=0.68\pm0.03$
\cite{key-14}. Hence, the experiment \cite{key-14} confirmed that
the coefficient $k$ in Eq. (\ref{eq: k}) substantially exceeds $0.5.$
One can find the scheme of this Mössbauer rotor experiment in Figure
1, see \cite{key-14} for details. 

The puzzle has been solved by us in {[}4\textendash 8{]}. The key
point is that in previous works in the literature {[}2, 3, 13\textendash 19{]}
an important effect of clock synchronization of has not been considered
for about 50 years, starting from the aforementioned book by Pauli
on relativity \cite{key-2}, up to the recent experiments \cite{key-13,key-14}. 

The Langevin transformation in cylindrical coordinates is about moving
from a Lorentz-Minkowski coordinate system, centered in the source
of the apparatus in Figure 1, having the $z-axis$ perpendicular to
the plane of the Figure, to a second coordinate system rotating around
the $z-axis$(Langevin frame). In the following the coordinates and
the proper time of the Lorentz-Minkowski frame will be labelled with
the subscript $L,$ while the coordinates and the proper time of the
observer in the Langevin frame will be labelled with the subscript
$R$ ($R$ stays for rotating). The Lorentz-Minkowski frame is the
laboratory frame, and the metric is {[}4\textendash 8, 20{]}

\begin{equation}
ds^{2}=c^{2}dt_{L}^{2}-dr_{L}^{2}-r_{L}^{2}d\phi_{L}^{2}-dz_{L}^{2}.\label{eq: Minkowskian}
\end{equation}
The Langevin transformation generates the Langevin reference frame
$\left\{ t_{R},r_{R},\phi_{R},z_{R}\right\} ,$ having constant angular
velocity $\omega$ around the $z-axis.$ In details, one sets {[}4\textendash 8,
20{]}

\begin{equation}
\begin{array}{cccc}
t_{L}=t_{R}\; & r_{L}=r_{R} & \;\phi_{L}=\phi_{R}+\omega t_{R}\quad & z_{L}=z_{R}\end{array}.\label{eq: trasformazione Langevin}
\end{equation}
This leads to Langevin metric for the Langevin frame {[}4\textendash 8,
20{]}
\begin{equation}
ds^{2}=\left(1-\frac{r_{R}^{2}\omega^{2}}{c^{2}}\right)c^{2}dt_{R}^{2}-2\omega r_{R}^{2}d\phi_{R}dt_{R}-dr_{R}^{2}-r_{R}^{2}d\phi_{R}^{2}-dz_{R}^{2}.\label{eq: Langevin metric}
\end{equation}
Considering the EEP, the line element (\ref{eq: Langevin metric})
is interpreted in the Einsteinian sense in terms of a stationary gravitational
field {[}4\textendash 8, 20{]}. Technical details of the last sentence
can be found in paragraph 89 of \cite{key-20}. The EEP means that
the inertial force experienced by a Langevin observer is interpreted
in terms of a gravitational ``force'' {[}4\textendash 8{]}. The
first well known effect arises from the ``gravitational redshift''
{[}4\textendash 8{]}. This effect is well known in previous literature
{[}3\textendash 10, 12-19{]}. Kündig's standard interpretation \cite{key-3}
argues that the Langevin observer concludes that his clock is slowed
down by the ``gravitational potential''. Hence, the clock of the
observer in the laboratory must be faster than the clock of the Langevin
observer. This is the so-called gravitational redshift``. Then, the
fractional energy shift in the laboratory results {[}3\textendash 10,
12-19{]}
\begin{equation}
\frac{E_{2}-E_{1}}{E_{1}}=\frac{\triangle E_{1}}{E_{1}}=\simeq-\frac{1}{2}\frac{v^{2}}{c^{2}}.\label{eq: fractional energy shift}
\end{equation}
Hence, one gets $k_{1}=\frac{1}{2}$ as the contribution to $k$ from
the first effect {[}3\textendash 10, 12-19{]}. 

The key point for needing an additional effect is the following. The
computation that permits to obtain Eq. (\ref{eq: fractional energy shift})
is usually performed in the Langevin reference frame. {[}4\textendash 8{]}.
But the final detector is in motion with respect to the Langevin observer
{[}4\textendash 8{]}, see Figure 1. Thus, the clock of the fixed observer
in the Lorentz-Minkowski laboratory frame is \emph{not} synchronized
with the clock of the Langevin observer. This implies that an additional
effect must contribute to the total effect {[}4\textendash 8{]}. This
effect of clock synchronization was not considered in previous literature
on the subject {[}2, 3, 13\textendash 19{]}. The general relativistic
approach in \cite{key-7,key-8} permitted us to explain this second
effect in a complete way. Here we show that this additional effect
can be obtained in another way via the third postulate of relativity
\cite{key-21}, which states that, if one considers a physical system
(the Langevin observer in the current case) moving through a flat
space-time, than there is at any moment a local inertial system such
that, in it, the system is at rest (the Langevin reference frame is
locally inertial). In that case, at any instant, the coordinates and
state of the system can be Lorentz transformed to the other system
through some Lorentz transformation. This means that if one considers
the frame of reference of the rotating Langevin observer, where the
observer is at rest, the coordinates and state of the system can be
Lorentz transformed to the Lorentz-Minkowski frame of reference of
the laboratory, which moves with respect to the frame of reference
of the Langevin observer. Relative to the Lorentz-Minkowski frame
of reference of the laboratory, the instantaneous velocity of the
rotating Langevin observer is $\vec{v}(t_{L})$ with magnitude $\left|\vec{v}\right|=v$
bounded by the speed of light $c,$ so that $0\leq v<c.$ Here the
time $t_{L}$ is the time as measured in the Lorentz-Minkowski frame
of reference of the laboratory, which corresponds with the proper
time of the laboratory $\tau_{L,}$ it is not the time measured in
the frame of reference of the rotating Langevin observer. The rotating
Langevin frame of reference and the Lorentz-Minkowski frame of reference
of the laboratory are Lorentz connected via the instantaneous Lorentz
factor \cite{key-21}
\begin{equation}
\gamma\equiv\frac{1}{\sqrt{1-\frac{v^{2}}{c^{2}}}}.\label{eq: istantaneous Lorentz factor}
\end{equation}
Then, one gets via the definition of the instantaneous Lorentz factor
\begin{equation}
\frac{d\tau_{L}}{d\tau_{R}}=\gamma,\label{eq: tempi infinitesimi}
\end{equation}
where $\tau_{R}$ is the proper time measured in the frame of reference
of the rotating Langevin observer. Hence, there is a time dilation
between the two frames of references given by
\begin{equation}
d\tau_{L}=\gamma d\tau_{R},\label{eq: tempi finiti}
\end{equation}
which means that the proper time between two ticks as measured in
the frame in which the clock is moving (the Lorentz-Minkowski frame
of reference of the laboratory), is longer than the proper time between
these ticks as measured in the rest frame of the clock (the frame
of reference of the rotating Langevin observer). In order to compute
$\gamma$ one must compute $v.$ For a radial coordinate $r_{L}$
it is obviously $v=\omega r.$ Thus, one gets: 
\begin{equation}
d\tau_{R}=\frac{d\tau_{L}}{\gamma}=\sqrt{1-\frac{v^{2}}{c^{2}}}d\tau_{L}=\sqrt{1-\frac{\omega^{2}r_{L}^{2}}{c^{2}}}d\tau_{L},\label{eq: gamma}
\end{equation}
which means that the clock of the rotating Langevin observer and the
clock of the Lorentz-Minkowski observer in the laboratory measure
the same proper time \textbf{if and only if} $r_{L}=r_{R}=0,$ i.e.
only in the origin of the two reference frames. Thus, clock synchronization
is necessary in all the other points. Hence, the contribution must
be calculated along the whole trajectory of photons. Thus, one must
find $r_{L}$ as a function of the time $\tau_{L}$ in Eq. (\ref{eq: gamma})
for the light rays. Clearly, as $r_{L}$ and $\tau_{L}$ refer to
the Lorentz-Minkowski frame of reference of the laboratory it is obviously
$r_{L}=c\tau_{L}.$ Then, Eq. (\ref{eq: gamma}) becomes

\begin{equation}
d\tau_{R}=\sqrt{1-\omega^{2}\tau_{L}^{2}}d\tau_{L}.\label{eq: secondo contributo finale}
\end{equation}
One can approximate Eq. (\ref{eq: secondo contributo finale}) with
\cite{key-7}
\begin{equation}
d\tau_{R}\simeq\left(1-\frac{1}{2}\omega^{2}\tau_{L}^{2}+....\right)d\tau_{L}.\label{eq: well approximated}
\end{equation}
Eq. (\ref{eq: well approximated}) takes into account the second effect
of order $\frac{v^{2}}{c^{2}}$ to time dilation, which, integrated
along the entire trajectory of the photons, gives 
\begin{equation}
\Delta\tau_{RT}=\int_{0}^{\tau_{LT}}\left(1-\frac{1}{2}\omega^{2}\tau_{L}^{2}\right)d\tau_{L}-\tau_{LT}=-\frac{1}{6}\tau_{L}^{3}\omega^{2}=-\frac{1}{6}r_{LT}\frac{v^{2}}{c^{2}}.\label{eq: delta tau 2}
\end{equation}
The subscript $T$ in the quantities in Eq. (\ref{eq: delta tau 2})
stands for \textquotedbl trajectory``. $\Delta\tau_{RT}$ represents
the difference between the proper time that has been measured by the
rotating Langevin observer and the proper time that has been measured
by the fixed Lorentz-Minkowski observer and $r_{LT}$ is the radial
coordinate of the detector. Then, the additional effect of clock synchronization
(at order $\frac{v^{2}}{c^{2}}$) to the energy shift is 
\begin{equation}
z\equiv\frac{\Delta\tau_{RT}}{\tau_{LT}}=\frac{\triangle E_{2}}{E_{1}}=-k_{2}\frac{v}{c^{2}}^{2}=-\frac{1}{6}\frac{v^{2}}{c^{2}}.\label{eq: z2}
\end{equation}
This means $k_{2}=\frac{1}{6}$.

Therefore, Eqs. (\ref{eq: fractional energy shift}) and (\ref{eq: z2})
permit to obtain the total fractional energy shift as 
\begin{equation}
\frac{\triangle E}{E_{1}}=\frac{\triangle E_{1}}{E_{1}}+\frac{\triangle E_{2}}{E_{1}}\simeq-\frac{2}{3}\frac{v^{2}}{c^{2}}.\label{eq: z totale}
\end{equation}
This theoretical result is completely consistent with the experimental
result $k=0.68\pm0.03$ in \cite{key-14} and with previous analyses
in {[}4\textendash 10, 22{]}, but in this letter it has been obtained
via a new, different analysis, which uses the third postulate of relativity.

Here are some additional considerations. After our original discovery
of the additional effect of clock synchronization in \cite{key-4},
this result has been independently confirmed by other Authors \cite{key-9,key-10,key-22}.
The experimental group that reanalysed the original data of Kündig
\cite{key-13} and realized the new experiment \cite{key-14} also
realized a second experiment with some additional collaborators \cite{key-23}.
In this new experiment they found the result $k=0.69\pm0.02$ \cite{key-23},
which is still consistent with the theoretical prediction of Eq. (\ref{eq: z totale}).
For the sake of completeness, we stress the important analogy between
the effect discussed here and the use of relativity theory in Global
Positioning Systems (GPS) \cite{key-24}. The additional term $-\frac{1}{6}$
in Eq. (\ref{eq: z2}) is similar to the correction that one has to
consider in GPS when one accounts for the difference between the time
measured in a frame co-rotating with the Earth geoid and the time
measured in a non-rotating Earth centered frame, which is locally
inertial {[}4\textendash 8{]} (and also the difference between the
proper time of an observer at the surface of the Earth and at infinity).
In fact, by simply considering the redshift due to the Earth's gravitational
field, but neglecting the effect of the Earth's rotation, GPS cannot
work, see {[}4\textendash 8{]} for details. 

Finally, the attentive reader may be interested in some other useful
papers regarding the topic covered in this letter, which are Refs.
{[}25\textendash 27{]}.

\subsection*{Conclusion remarks}

Based on its renewed interest, due to the discovery and explanation
of an additional effect of clock synchronization which has been missed
for about 50 years, in this letter we addressed the theoretical explanation
of such an additional effect via a novel approach founded on the third
postulate of relativity by finding complete consistence with the experimental
results in \cite{key-14,key-23} and with previuos theoretical approaches
in {[}4\textendash 10, 22{]}.

\subsection*{Data availability statement}

This manuscript has no associated data.

\end{document}